\documentclass{evn2024}
\usepackage{graphicx}
\usepackage{natbib}

\defcitealias{itu_p452_16}{Rec.~ITU-R~P.452-16}
\defcitealias{itu_p452_17}{Rec.~ITU-R~P.452-17}
\defcitealias{itu_p452_18}{Rec.~ITU-R~P.452-18}
\defcitealias{itu_p2001_5}{Rec.~ITU-R~P.2001-5}
\defcitealias{itu_p2108_1}{Rec.~ITU-R~P.2108-1}
\defcitealias{itu_p2109_2}{Rec.~ITU-R~P.2109-2}
\defcitealias{itu_p676_13}{Rec.~ITU-R~P.676-13}
\defcitealias{itu_m1583_1}{Rec.~ITU-R~M.1583-1}
\defcitealias{itu_p1144_9}{Rec.~ITU-R~P.1144-9}
\defcitealias{itu_ra769_2}{Rec.~ITU-R~RA.769-2}
\defcitealias{itu_ra1513_2}{Rec.~ITU-R~RA.1513-2}
\defcitealias{itu_ra1631_0}{Rec.~ITU-R~RA.1631-0}
\defcitealias{itu_f1766_0}{Rec.~ITU-R~F.1766-0}
\defcitealias{itu_s1586_1}{Rec.~ITU-R~S.1586-1}
\defcitealias{itu_ra2332}{Rep.~ITU-R~RA.2332}
\defcitealias{itu_ra2507}{Rep.~ITU-R~RA.2507}
\defcitealias{itu_res681}{ITU-R~Resolution~681}
\defcitealias{cispr11}{EN\,550011 (CISPR-11)}
\defcitealias{ecc_report_271}{ECC~Report~271}
\defcitealias{ecc_report_308}{ECC~Report~308}
\defcitealias{ecc_report_348}{ECC~Report~348}
\defcitealias{ecc_report_327}{ECC~Report~327}
\defcitealias{ecc_report_350}{ECC~Report~350}
\defcitealias{ecc_report_351}{ECC~Report~351}
\defcitealias{ECSS-E-ST-20-07C}{ECSS-E-ST-20-07C}
\defcitealias{MSFC-SPEC-521C}{MSFC-SPEC-521C}
\defcitealias{radioregs}{Radio Regulations}
\defcitealias{un_resolution_ga69-266}{Resolution GA 69/266}

\begin{document}
   \title{Spectrum management and the EVN}

   \author{
       B. Winkel\inst{1,4}\thanks{On behalf of the Committee on Radio Astronomy Frequencies (CRAF)}
       \and
       F. Giovanardi\inst{2,4}
       \and
       M. Lindqvist\inst{3,4}
       }

   \institute{
         Max-Planck-Institut f\"ur Radioastronomie, Auf dem H\"ugel 69, 53121 Bonn, Germany
         \and
         Arcetri Astrophysical Observatory INAF, Largo Enrico Fermi 5, Florence, Italy
         \and
         Department of Space, Earth and Environment, Chalmers University of Technology, Onsala Space Observatory, 439 92 Onsala, Sweden
         \and
         European Science Foundation, Committee on Radio Astronomy Frequencies, 1, quai Lezay Marnésia BP 90015, F-67080 Strasbourg Cedex, France
         }

   \abstract{
      In recent years, the utilisation of the radio spectrum has dramatically increased. Digital telecommunication applications, be it terrestrial cell-phone networks or new-space low-earth orbit satellite constellations, have not only acquired unprecedented  amounts of spectrum but also use their frequencies everywhere on Earth. The consequences for radio astronomy and other scientific radio services are severe. A single cell-phone tower within hundreds of kilometers around a radio telescope can blind us and there is no place on Earth to escape the ubiquitous transmissions of satellite megaconstellations.

      Since 1988, the Committee on Radio Astronomy Frequencies (CRAF) is advocating for astronomers' rights to use the spectrum. CRAF does this by participation in the national and international regulatory frameworks. Hundreds if not thousands of documents need to be processed every year. CRAF not only contributes to regulatory texts, but even more importantly, performs spectrum compatibility calculations. In this contribution, CRAF's latest activities are summarized with a focus on matters relevant to EVN operations.
   }

   \maketitle
%

\section{Introduction}

The radio spectrum is a scarce resource. Since the early days of radio astronomy, human-made radio signals were a potential source of interference for the highly sensitive obversions of the universe. However, in recent years, the usage of the spectrum has dramatically increased and spectrum efficiency of the active radio services continues to grow, leaving less and less room for radio astronomers to use undisturbed portions of the spectrum.

To mitigate this, observatories were typically built in remote areas, ideally with natural terrain shielding to improve the situation. While this helped to some extent, powerful transmitters such as radars or cell-phone base stations can outshine the faint radio waves from the universe by orders of magnitude, even over great distances. Additionally, the widespread deployment of satellite communication systems in low Earth orbit makes it difficult to avoid human-made signals, even at the most remote sites.

The importance of radio astronomy was recognised many decades ago, and protective measures were implemented by the Radiocommunication Sector of the International Telecommunication Union (ITU-R). In the so-called \citetalias{radioregs} (RR) of the ITU-R, which has the status of an international treaty, the radio astronomy service (RAS) is established as one of the passive services for scientific purposes. There are several frequency bands, which are allocated to the RAS on a primary or secondary basis. Although the total amount of the spectrum allocated to the RAS (especially below 100 GHz) is too small to satisfy the needs of modern astronomy, the protection of these frequency intervals already requires enormous efforts from radio astronomy spectrum managers, simply because the economic pressure has increased so much.

The Committee on Radio Astronomy Frequencies (CRAF) of the European Science Foundation (ESF) is a collaboration of spectrum managers from European radio astronomy observatories and research facilities. The aim of CRAF is to advocate for the protection of passive radio service operations, with a focus on radio astronomy and geodetic VLBI. Details on the history of CRAF and its organisational structure can be found in a recent report \citep{craf23}. CRAF also represents the majority of the EVN observatories in spectrum management matters.

This paper provides a summary of CRAF's activities in spectrum management, focusing on matters relevant to EVN operations. In Section~\ref{sec:regframework} an overview of the regulatory framework is provided. Section~\ref{sec:craf} explains how CRAF operates and what its work entails. Finally, in Section~\ref{sec:topics}, a non-exhaustive list of current topics is presented, which may be of interest to the EVN community.

\section{Regulatory framework}\label{sec:regframework}
Astronomers have a tendency to label any anthropogenic signal in their data as radio frequency interference (RFI). From a regulatory perspective, however, this is far from the truth. Active radio services have rights to use the spectrum, too, and the allocations and licences or operation permits, which are required, are usually the result of long processes between and within administrations involving many stakeholders. Therefore, active use of the spectrum use can be considered as agreed upon by our society. In legal terms, RFI refers to any kind of unwanted signal, e.g. spectral side-lobes of a signal or harmonics, that enters the system of another radio service on frequencies that are allocated to that service. However, in some cases, when two services share a frequency band, it could also mean that wanted signals are affecting another system.

As radio waves do not stop at national borders, there is a clear need for an international framework for coordinating radio services. The ITU-R, a specialized agency of the United Nations, is the main body for establishing and developing this framework. ITU-R member countries can update the radio regulations and other regulatory documents (resolutions, recommendations, and reports), but in almost all cases full consensus is required. This means that processes can take a long time and often require a lot of compromises. The radio regulations as the main regulatory document can only be amended at world radiocommunication conferences (WRCs), which are held every three to four years. Most of the work on recommendations and reports, as well as the preparation of WRCs, is done in so-called study groups. Non-governmental stakeholders such as CRAF are called sector members and are invited to participate in all these groups and in the WRC, but they do not have the right to vote.

At the core of the RR, there is the table of frequency allocations (TFA). It lists the allocations to specific radio services for each block of the radio spectrum from 8.3 kHz up to 275 GHz. Frequencies above 275 GHz and below 1 THz are not currently allocated to services, but a list with identifications is provided. The TFA also contains many footnotes with additional rules. In some cases there are specific regional or country allocations, in other cases they contain protection criteria or other information needed for coordination. The ITU-R also maintains a list of radio stations in the master international frequency register (MIFR), which is an important prerequisite for requesting protection from interference, especially from other countries or satellites, which cannot be controlled by the national administration of the affected station. Moreover, all RAS observatories should be notified to the MIFR.

Allocations can be of primary or secondary nature. Primary and secondary allocations define the priority of the service in relation to other services that may share a part of the spectrum. Therefore, only with a secondary allocation, a service cannot ask for protection from a primary service in the same band, and it must not interfere with the primary service. RAS has several primary and secondary allocations. Below 4~GHz, only 5\% are reserved for RAS (1.5\% in primary allocations). Some bands are also subject to the footnote RR\,5.340, which prohibits emissions completely. This was introduced for the passive services, such as the Earth exploration satellite service (EESS) and the RAS, to protect the most important frequencies, e.g., the band 1400$-$1427~MHz. Another noteworthy footnote is RR\,5.149, which ``urges administrations to take all practicable steps to protect the radio astronomy service from harmful interference'' in a number of bands during the assignment process. Assignment refers to granting national licences or operating permits, in contrast to the ITU-R allocations. Thus, even if an administration is willing to protect a particular RAS site on the basis of this footnote, there is little legal obligation for neighbouring countries to do the same.

\begin{table*}
   \caption[]{Examples of ITU-R recommendations and reports that are relevant for radio astronomy.}
      \label{tab:ras-recommendations}
  $$
      \begin{array}{p{0.23\linewidth}l}
         \hline
         \noalign{\smallskip}
         Document      &  \textrm{Topic} \\
         \noalign{\smallskip}
         \hline
         \noalign{\smallskip}
         Recs. M.1583 \& S.1586 & \textrm{Compatibility calculation methods for non-GSO satellite constellations}\\
         Rec. M.2101 & \textrm{5G cell phone networks}\\
         Recs. P.452 \& P.2001 & \textrm{Path propagation loss models (terrestrial)}\\
         Rec. P.676 & \textrm{Atmospheric attenuation}\\
         Recs. P.2108 \& P.2109 & \textrm{Clutter and building entry losses}\\
         Rec. RA.314 & \textrm{Preferred frequency bands for RAS}\\
         Rec. RA.517 & \textrm{Protection from out-of-band emissions}\\
         Recs. RA.611 \& SM.329 & \textrm{Protection from spurious emissions}\\
         Rec. RA.769 & \textrm{Protection criteria for RAS}\\
         Rec. RA.1513 & \textrm{Acceptable levels of data loss for RAS}\\
         Recs. RA.1631 \& SA.509$^{\mathrm{a}}$ & \textrm{Reference antenna pattern for RAS}\\
         Rec. RS.2066 & \textrm{Protection from strong SAR satellites in X-band}\\
         Rec. S.1528 & \textrm{Satellite antenna patterns ($<$30 GHz)}\\
         Rec. SM.329 & \textrm{Unwanted emissions in the spurious domain}\\
         Rec. SM.1541 & \textrm{Unwanted emissions in the out-of-band domain}\\
         Rec. SM.1542 & \textrm{Protection from unwanted emissions}\\
         Rep. RA.2131 & \textrm{Supplementary information on the detrimental threshold levels of interference}\\
         Rep. RA.2188 & \textrm{Power levels potentially damaging RAS receivers}\\
         Rep. RA.2259 & \textrm{Radio Quiet Zones}\\
         Rep. RA.2507 & \textrm{Geodetic VLBI}\\
         \noalign{\smallskip}
         \hline
      \end{array}
  $$
\begin{list}{}{}
\item[$^{\mathrm{a}}$] Rec. SA.509 is based on very old data, but is still referenced in some places.
\end{list}
\end{table*}

In Tab.~\ref{tab:ras-recommendations} a number of ITU-R recommendations and reports are listed, which are relevant to the protection of RAS\footnote{All recommendations and reports of the ITU-R are publicly available at {\tt https://www.itu.int/pub/R-REC} and {\tt https://www.itu.int/pub/R-REP}.}. A detailed discussion of all of them is beyond the scope of this text, but a few of them deserve a brief introduction. The most fundamental recommendation is \citetalias{itu_ra769_2}, which describes the protection criteria for RAS. For single-dish operation, these are based on the noise levels that can be achieved after a given integration time in a given bandwidth: an interfering signal must not contribute more power to the receiver than 10\% of the RMS noise level assuming state-of-the-art receiver noise temperatures and typical antenna temperatures (e.g., from ground or atmospheric contributions). Although RA.769 does not specify the integration time to be used for the calculation, the example tables 1 and 2 in the recommendation show the values for 2000 seconds. This has become a de facto standard in spectrum management calculations and most administrations will not accept any deviation from this integration time. A distinction is made between continuum and spectral line observations. For the latter, the bandwidth is much smaller, of the order of typical channel widths of RAS spectrometers. VLBI thresholds are determined differently. Here, the contribution of any interfering signal must remain below 1\% of the system temperature of the receiver system. The VLBI thresholds are less stringent, which makes sense as VLBI is less affected by (local) interference.

For compatibility calculations, the path propagation loss, i.e., the attenuation of the transmitted signal on the path between transmitter and receiver, is important. There is an entire study group, SG\,3, at ITU-R, which develops appropriate models. Table~\ref{tab:ras-recommendations} lists a few of them. Often, \citetalias{itu_p452_18} and \citetalias{itu_p2001_5} are used for terrestrial propagation. At higher frequencies, the atmospheric attenuation becomes a relevant factor, too, which is the subject of \citetalias{itu_p676_13}. Clutter and building entry losses are contained in \citetalias{itu_p2108_1} and \citetalias{itu_p2109_2}. It is important to point out that while these models are often inferior to specialised state-of-the-art scientific models (e.g., for atmospheric attenuation), ITU-R usually accepts only calculations that are made with their models. This has the benefit that consensus among all member countries was reached on the details of a model and will prevent lengthy discussions.

The ITU-R is responsible for the international framework, but there are also several regional spectrum management groups, such as the African Telecommunication Union (ATU), the Asia-Pacific Telecommunity (APT), or the European Conference of Postal and Telecommunications Administrations (CEPT). For the EVN observatories, the most relevant of these groups will be CEPT, and CRAF participates in its activities. CEPT not only prepares the WRCs with regulatory and technical studies aimed at reaching a common European position on the agenda items, but it also runs many technical and regulatory working groups for purely European issues. As Europe is densely populated, with many countries in a relatively small area, a high degree  of multilateral coordination is required. In many cases, European countries are seeking to harmonise their spectrum licensing regimes, in particular within the framework of the European Union.

CRAF is also involved in national activities. Most European administrations have internal preparatory groups in which the various local stakeholders try to reach a common position which the administration would then present at CEPT or ITU-R. As administrations naturally put more effort into protecting observatories on their own territory, it is very important for radio astronomers to establish good relations with their national administration.

\begin{figure*}
   \centering
   \includegraphics[bb=50 70 920 420,width=0.9\textwidth,clip]{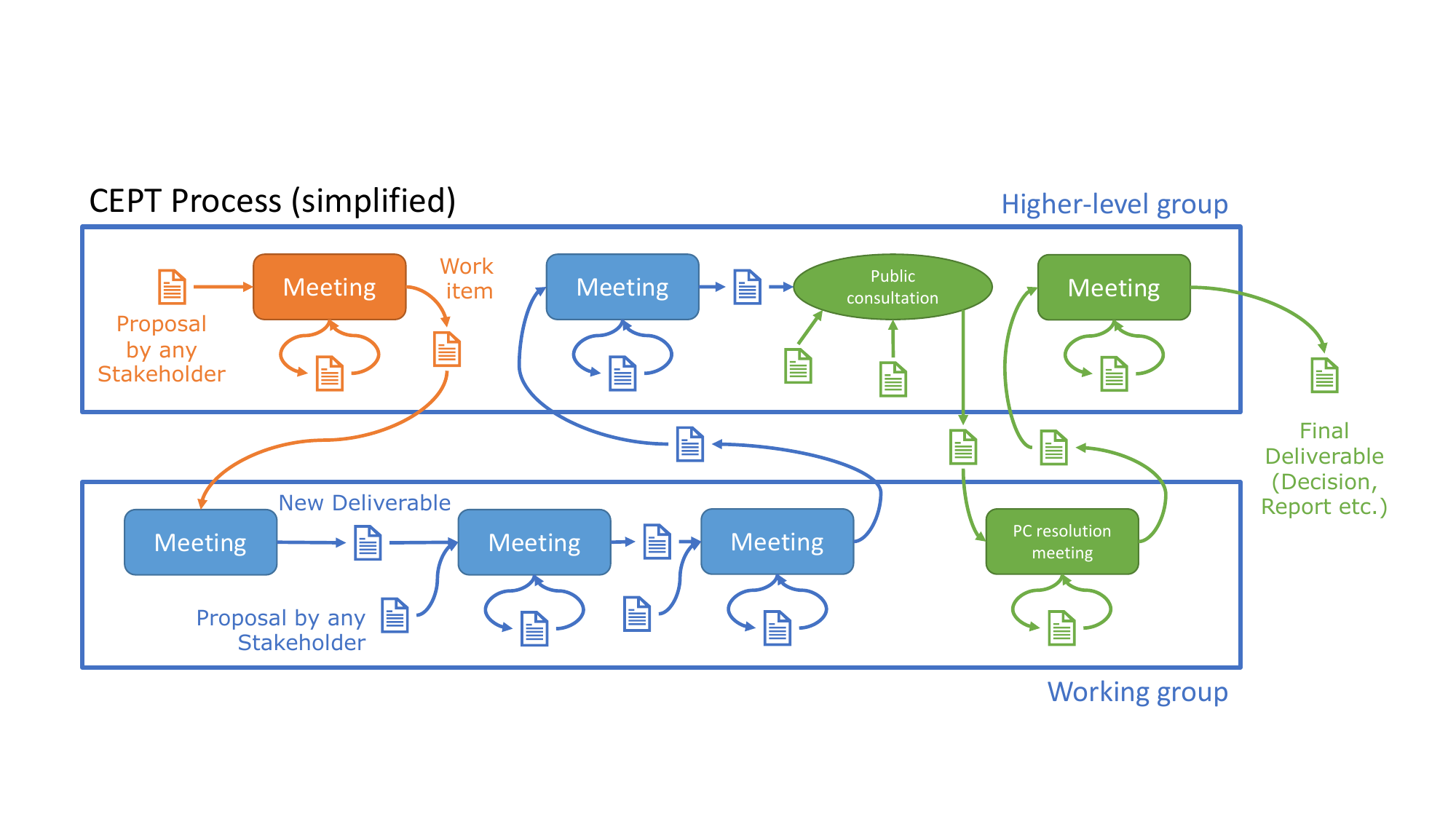}
   \caption{Input-document driven process in CEPT.}\label{fig:cept-process}
\end{figure*}

In almost all the cases of participation in spectrum management organisations, the underlying work method is known as ``input document driven''. That means, without the timely submission of an input document before the deadline, a topic will not be discussed. The process is shown in Fig.~\ref{fig:cept-process} for the example of the CEPT. Furthermore, most of the work is organised into so-called work items (WI), which have a well-defined scope, deadlines and deliverables (e.g., a new report or recommendation). While stakeholders or administrations are free to submit any document to a meeting, the groups will prioritize the WI-related documents. Thus, if any party wishes to work towards a new deliverable, they first need to make a proposal to one of the high-level groups, which then discuss the matter and upon agreement will create a formal WI. This WI is then assigned to one of the technical or regulatory subgroups, which will carry out the actual drafting work. Anyone can contribute via input documents to the work group meetings. As soon as the work is finished, the draft will be sent to the higher-level group again, which will decide if the document is ready for the next step, the public consultation phase. This takes a few months and any interested party can submit comments. It should be noted though, that administrations and working groups will usually not be happy, if huge amounts of new contents are submitted in this late phase. The subgroup will incorporate all comments received into the document and again send it to the higher-level group, which will then adopt it as appropriate. While this process is slightly different in the ITU-R, the general idea is the same.

It is very important to understand these processes and follow them closely. Procedural mistakes can mean that inputs are not considered. Furthermore, even if one was successful in incorporating a section or paragraph into a document, one must follow the process until the end, as another party could propose to delete that text at a later meeting. Although all stakeholders are invited to contribute to the meetings, the decision-making is up to administrations, only. Thus, if disagreements on a contribution are anticipated, it is advisable to advocate for support from national administrations in advance. Furthermore, one should be prepared to compromise, as the documents can only be adopted if all administrations agree.

\section{The Committee on Radio Astronomy Frequencies (CRAF)}\label{sec:craf}
The complex framework and working principles described in the previous section make it is very difficult for individuals to successfully participate in the meetings. There is a huge number of relevant documents and meetings to attend. For this reason, the Committee on Radio Astronomy Frequencies was established in 1988. Since then, CRAF has been involved in numerous activities within ITU-R and CEPT. Technically, CRAF is a so-called expert committee of the European Science Foundation. It has no individual members, but member institutes that propose delegates to the committee. Most of the European radio astronomy observatories are represented. The International VLBI Service for Geodesy and Astrometry (IVS), as well as the Institut de radioastronomie millim\'{e}trique (IRAM) are also members. In addition, the Square Kilometre Array Observatory (SKAO), the European Southern Observatory (ESO), and the European Space Agency (ESA) are observer members of CRAF. CRAF has a service contract with JIVE, which provides support for certain spectrum management tasks.

It should not be unmentioned that there are also several sister committees to CRAF in other ITU-R regions. In the US, the Committee on Radio Frequencies (CORF) of the National Academies of Sciences, Engineering, and Medicine advocates for passive use of the radio spectrum. There is also the Radio Astronomy Frequency Committee in the Asia-Pacific region (RAFCAP). Last but not least, on the global level, the Scientific Committee on Frequency Allocations for Radio Astronomy and Space Science (IUCAF) is active. CRAF cooperates closely with these international partners, especially at the ITU-R. CRAF also supports other scientific initiatives, e.g., to improve the protection of space weather observations at the ITU-R level or to preserve potential scientific sites on the Moon for future projects.

\begin{figure*}
   \centering
   \includegraphics[bb=10 0 450 340,width=0.45\textwidth,clip]{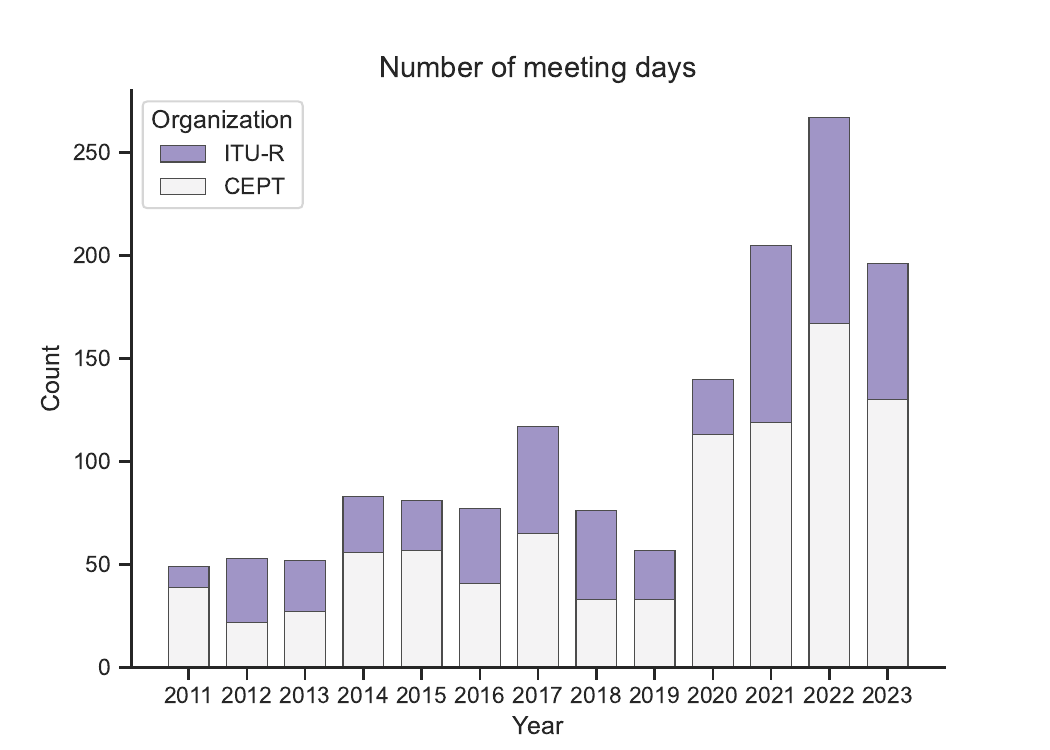}~
   \includegraphics[bb=10 0 450 340,width=0.45\textwidth,clip]{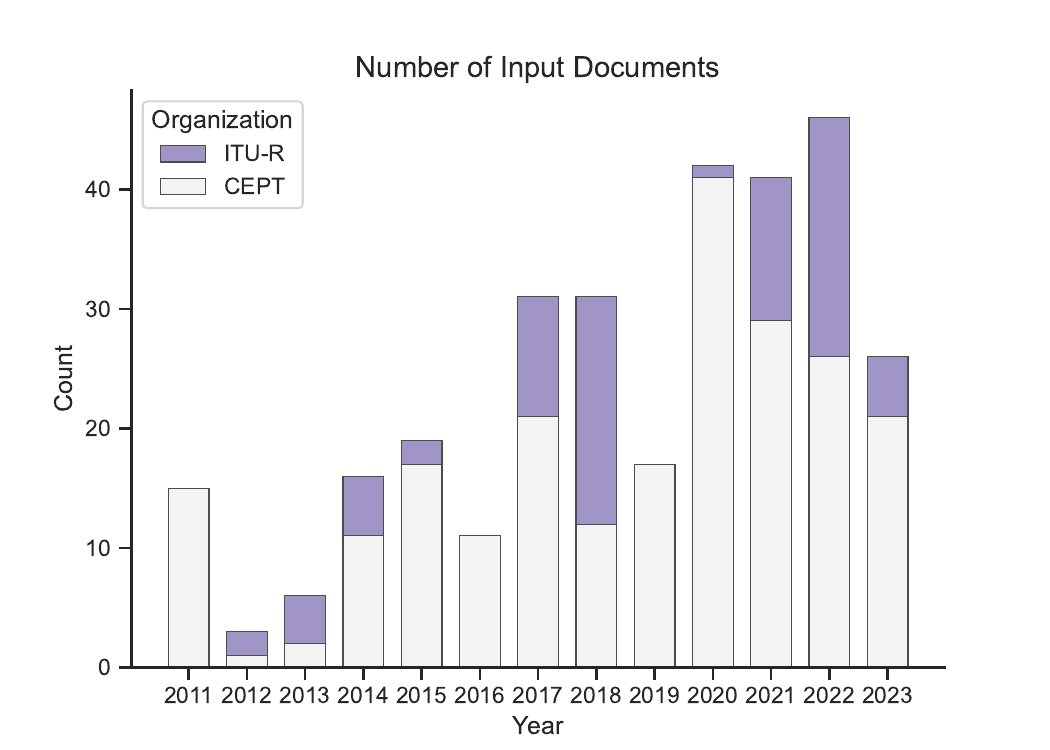}
   \caption{\textit{Left:} Number of meeting days with CRAF participation. \textit{Right:} Number of input documents submitted by CRAF.}\label{fig:craf-stats}
\end{figure*}

The workload of CRAF members has skyrocketed over the years, as reflected in the number of meetings attended and contributions made; see Fig.~\ref{fig:craf-stats}. Since the Covid-19 pandemy, most meetings allow remote participation, which has helped CRAF to participate in more relevant meetings than before without  higher travel costs. On average, the number of input documents produced by CRAF members is between 30$-$40 in recent years. While some of them only contain smaller contributions, especially for regulatory documents, technical studies can be very labour intensive, comparable to scientific publications.

\begin{figure}
   \centering
   \includegraphics[width=\columnwidth,clip]{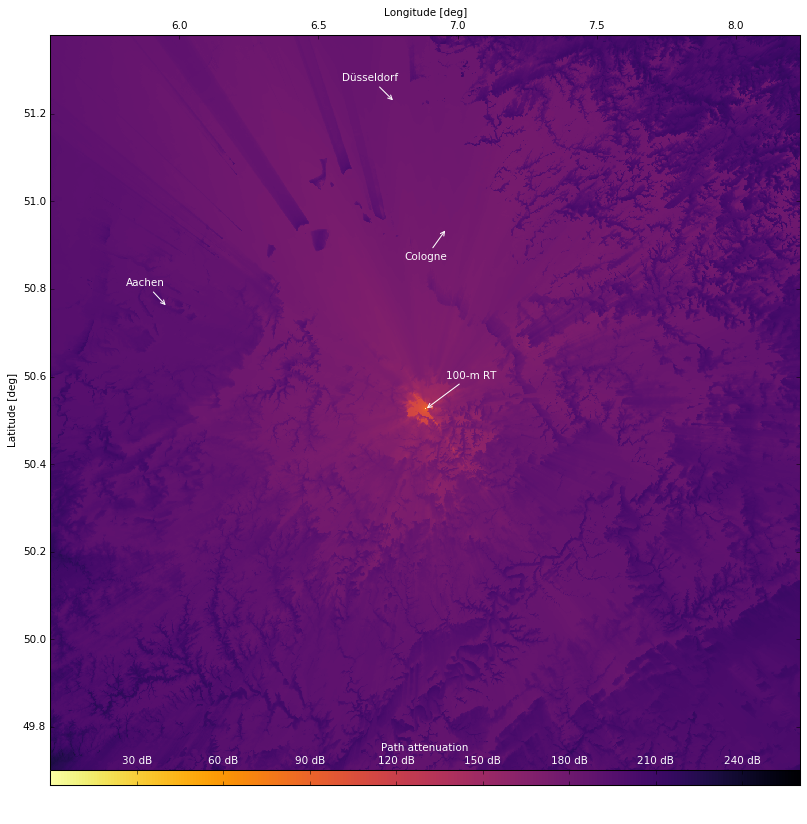}
   \caption{Example attenuation map visualising the path attenuation (in dB) for a frequency of 3.5~GHz around the 100-m telescope at Effelsberg, Germany. The colours indicate the propagation loss between each location in the map and the telescope in the centre. From CRAF self-evaluation report \citep{craf23}.}\label{fig:attenuation_map}
\end{figure}

CRAF has developed its own software tools for performing the spectrum compatibility studies, which have become widely used in the community already. There are even non-radio astronomers who like to work with them. The most mature CRAF software tool is \textit{pycraf} \citep{pycraf}, a package (library) for the Python programming language. It contains implementations of various path propagation models, antenna patterns, RAS threshold levels and much more. An example of an attenuation map is shown in Fig.~\ref{fig:attenuation_map}. The software is free and open source, and is hosted on GitHub\footnote{{\tt https://github.com/bwinkel/pycraf}}.

\section{Important spectrum management topics}\label{sec:topics}
In the following, a few of the current topics will be briefly presented, which are of relevance to the EVN community. Obviously, there are many activities in the radio spectrum these days, which have the potential to disturb astronomy observations of all kinds. VLBI is somewhat special, as it is less affected by local interference. However, if the input power to a VLBI station is too high, the sensitivity drops notably. The additional RFI power simply adds to the noise floor of the receiver. Very bright sources of anthropogenic signals are therefore most critical, usually belonging to long-distance communication signals (e.g. cell-phone towers, communication satellite constellations, digital TV broadcasting) or radar applications (e.g. weather or military radars). However, VLBI antennas need to be calibrated, which usually requires continuum observations in single-dish mode. Hence, low to medium power levels can also affect VLBI operations.

In light of the above, it is not surprising that most of the relevant activities are related to the ubiquitous deployment of new cell-phone towers on more and more frequencies, and the advent of so-called satellite mega-constellations -- communication satellites in low Earth orbit (LEO), which are deployed in huge numbers. CRAF is also working at the ITU-R level to improve the recognition of the VLBI for Geodesy (VGOS).

\subsection{Cell-phone networks}

\begin{figure*}
   \centering
   \includegraphics[width=0.9\textwidth,clip]{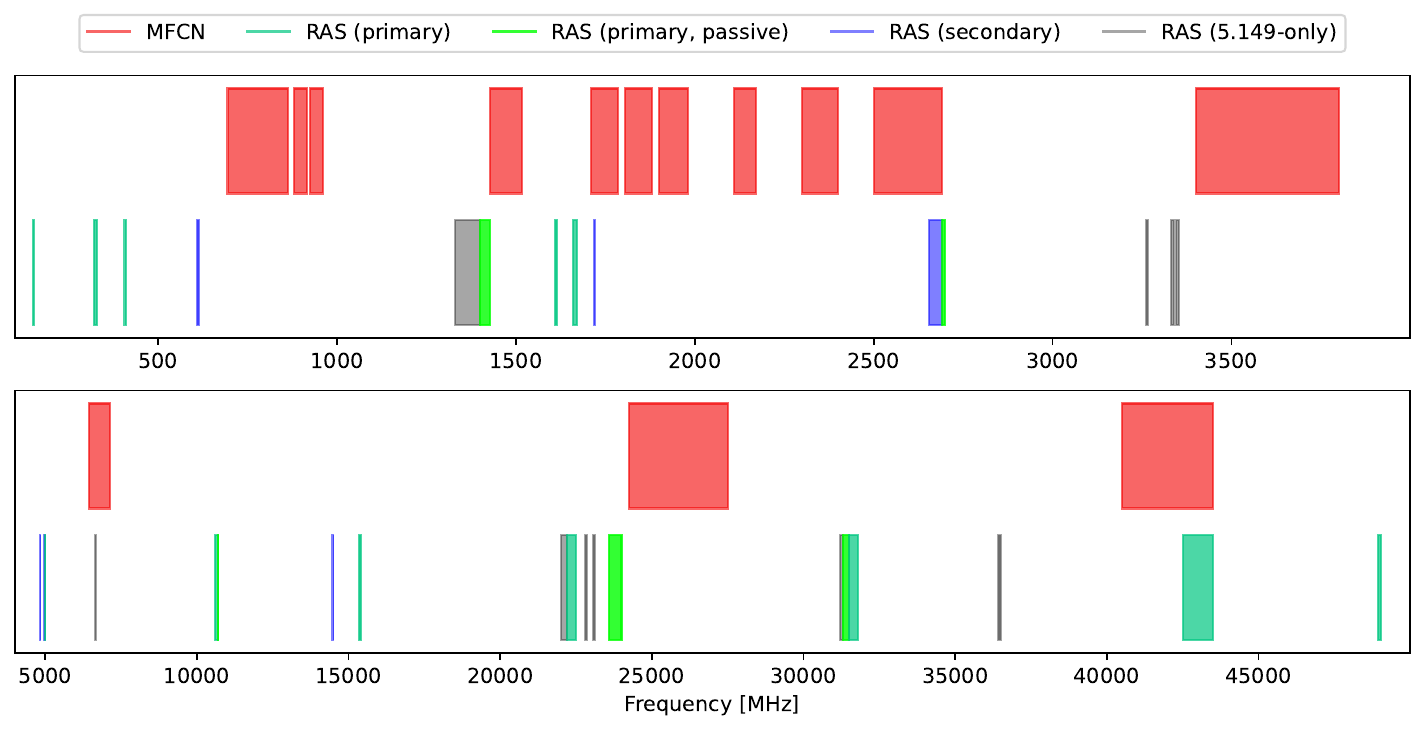}
   \caption{Comparison between the amount of spectrum reserved for cell-phone applications (MFCN) and radio astronomy.}\label{fig:mfcn}
\end{figure*}
Cell-phone networks (also known as Mobile-Fixed Communication Networks, MFCN), which are operated under the Mobile Service as the International Mobile Telecommunication (IMT) application at the ITU-R, are the single application that has received the largest amount of new spectrum over the past two decades. Below 4~GHz, IMT has identifications in the allocation table for a third of the spectrum. In Fig.~\ref{fig:mfcn} the situation is illustrated. The highest level of protection for RAS is only provided in the primary bands (green colour). In many European countries, administrations also try to coordinate with the RAS in bands with secondary allocation or 5.149-only protection (see Section~\ref{sec:regframework}). However, the total amount of spectrum reserved for RAS operations is tiny and certainly not sufficient for the needs of modern astronomy.

CRAF spent a lot of time performing compatibility calculations for the mm-wave 5G networks (above 24~GHz), that were studied in the 2015$-$2019 WRC cycle, as well as for the so-called mid-band at 6~GHz, where WRC-23 decided to identify several 100~MHz with IMT.\footnote{IMT is not a radio service itself. Instead allocations are made to the so-called Mobile Service, which is more general, and then certain frequency bands are `identified' for use with IMT applications. This is meant to streamline the use of cell-phone networks on a world-wide basis.} Use of IMT in the upper 6-GHz band will make observations of the important methanol spectral line a 6.67~GHz impossible, if no coordination measures are established. In Europe, CRAF is heavily advocating for a regional solution.

Compatibility calculations for IMT networks are among the most sophisticated studies, which CRAF is carring out. With the introduction of the 5G technology, active antennas are used on cell-phone towers forming antenna beams towards the end users in real time. This needs to be fully simulated to determine the side-lobe gain of the IMT antennas in the direction of the radio telescopes. The deployment of the networks (number and distribution of base stations around the observatories) plays a major role in the total aggregated power that could enter the RAS receivers. User terminals (e.g., smartphones) are also simulated, but usually contribute an order of magnitude less power, as they emit less and are closer to the ground and thus subject to clutter. Cell phones employ use power control algorithms in order to save energy when the link budget to the nearest base station allows it. At higher frequencies, user terminals could also feature beamforming. More details on the calculations can be found in e.g. \citetalias{ecc_report_308} and \citetalias{ecc_report_348}.

\subsection{New satellite constellations in low-Earth orbits}
It is estimated that the number of satellites in LEO could exceed 100 thousands by the end of the decade. Even today, we have more than ten thousand satellites in orbits, the majority of which were launched in the past few years. Two recent papers \citep{divruno23-starlink,bassa24}, have even shown that not only the communication signals (either in the wanted bands or via spectral sidelobes) can have an impact, but that on-board electrics and electronics can leak electromagnetic radiation, which can disturb low frequency observatories (e.g., LOFAR). To distinguish this leakage from the `classical' wanted and unwanted emissions of a satellite, which are subject to ITU-R regulation, the authors coined the term unintended electromagnetic radiation (UEMR). There is currently no sufficient regulation for UEMR from satellites.

While the UEMR issue has attracted considerable media and scientific interest, the `normal' spectrum management work is more important and binds more of CRAF's resources. Long before the topic of satellite megaconstellations hit the news, the satellite group of the CEPT/ECC was working on \citetalias{ecc_report_271}, which included also compatibility calculations for SpaceX/Starlink, OneWeb and several potential victim services including RAS. This report led to the protection of RAS stations in Europe, demonstrating that CRAF's work is key to successful coordination. Another milestone is the new ECC Report 363 (\textit{in public consultation}) on aggregated interference from multiple satellite constellations, with major contributions from CRAF members.

With more and more satellites in LEO, the interference potential obviously increases significantly. Although the status of RAS in the radio regulations is clear, there is still a risk that satellites will be launched, and it will only be discovered later that spectral sidelobes (including spurious emissions) are causing troubles. Once satellites are in orbit, interference cannot easily be fixed. To improve the coordination, CRAF started an initiative at ITU-R level which led to a new agenda item at WRC-23 \citepalias[see][]{itu_res681}, tasking ITU-R to study regulatory and technical measures that could improve the practical protection, ideally before launches. A second aspect of this agenda item is to study if and how radio quiet zones (RQZ) could be introduced at the international level. Today, existing RQZs are subject to national legislation and therefore cannot restrict satellite systems.

\subsection{VGOS}
Currently, VGOS stations are considered radio astronomy observatories, which ignores not only the very specific bandwidth requirements -- VGOS needs four larger blocks of spectrum in the range between 2 and 14~GHz -- but also the fact that the same frequencies need to be protected at \textit{every} station worldwide. If only some countries afford protection to their stations, this will not suffice to ensure the required VGOS data quality, which is requested by \citetalias{un_resolution_ga69-266} of the United Nations.

Therefore, CRAF started an initiative to improve the spectrum situation at the ITU-R level. The first step was to work on a new report \citepalias{itu_ra2507} and a new ITU-R recommendation is in drafting process. CRAF will also advocate for an agenda item at WRC-31, which could ask to study the feasibility of additional allocations.

Owing to the overlap between EVN and VGOS stations, CRAF's efforts on Geodetic VLBI had been previously presented in EVN fora \citep[e.g.][]{hase21,bautista23,divruno23-starlink}.

\subsection{Car radars}
As a final example, we will briefly discuss car radars. Not because these have more interference potential than other applications, but because it provides a nice demonstration of what has become possible with open source datasets. For car radars, it is usually better to perform aggregation studies, looking at the sum of the contributions from all cars in the area, rather than a worst-case scenario with a single car, simply because there can be many thousands of cars even in remote RAS locations.

Not only can the number of cars on specific road types be estimated using publicly available information, but by querying  OpenStreetMap\footnote{{\tt https://www.openstreetmap.org}} data, we were able to retrieve road networks around European sites. We used this to simulate cars in realistic locations and orientations. The latter is important when cars have multiple radars, e.g. front and corner radars. Furthermore, for the calculation of propagation losses, terrain heights are available from either SRTM \citep{farr07} or Lidar (laser ranging) measurements, and clutter information can be extracted from Corine\footnote{{\tt https://land.copernicus.eu/en/products/corine-land-cover}} land cover data \citep[see e.g.][]{feranec2016}.

\begin{figure}
   \centering
   \includegraphics[bb=5 5 640 615,width=\columnwidth,clip]{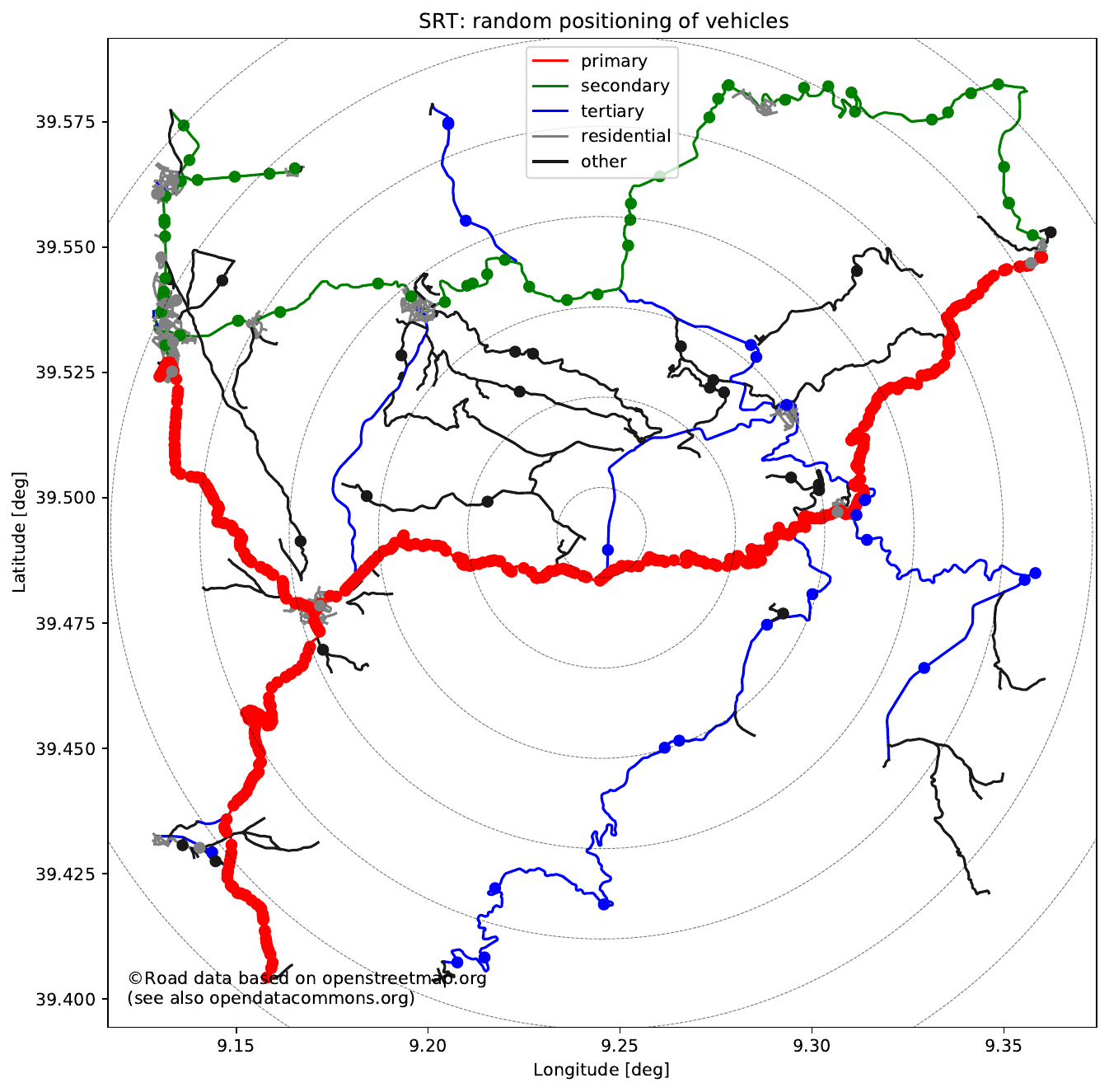}
   \caption{Road network data around the SRT used for simulations of car radars. Circles have radii of $1,\,3,\,5,\,\ldots$~km.}\label{fig:cars_srd}
\end{figure}

Fig.~\ref{fig:cars_srd} shows an example, where the car positions for a particular time step of the simulation are represented as filled circles over the road network in the vicinity of the Sardinia radio telescope (SRT). CRAF studies have contributed to e.g. \citetalias{ecc_report_327}, \citetalias{ecc_report_350}, and \citetalias{ecc_report_351}. An extension of the method is presented in \citet{giovanardi24}, where a hexagonal binning scheme is used to identify areas that contribute most to aggregated received power. This could be used as a basis for tailor-made coordination or exclusion zones.

\section{Summary and conclusions}\label{sec:conclusions}
This paper presented an overview of some of the recent activities of the Committee on Radio Astronomy Frequencies relevant to EVN operations. Although we were only able to provide a high-level report, it should have become clear that CRAF invests an enormous amount of work in the fight to keep the radio environment as free of interference as possible. CRAF members participate in a many national and international meetings and working groups, for which dozens of input documents are produced each year. One of CRAF's strengths is its expertise in carrying out compatibility calculations, even for the highly complex and computationally demanding issues we are faced with today.

However, the workload has been continuously  increasing over the years, and we urge the scientific community to invest more in spectrum management, both in terms of the financial contributions and human resources. Even today, CRAF members have to prioritise their work. The issue with spectrum management is that once decisions are made, they are hard to change. If access to a certain part of the spectrum is lost (e.g., because an active service is allowed to enter), the chances of getting it back are extremely slim. With our work, we think we have shown that a difference can be made, but it requires appropriate and continuous effort.

The EVN, in particular through JIVE, has a key role to play in supporting CRAF as it not only represents the European radio observatories, but also links two scientific fields, radio astronomy and geodesy. JIVE already provides services to CRAF and there is a strong collaboration. In addition, CRAF often needs feedback from the scientific communities, e.g., by quickly providing RFI reports when the spectrum situation changes, or by helping CRAF to collect good arguments why specific frequencies are important. An example of the latter was a joint effort between CRAF and members of the VLBI community involved in 6.6-GHz methanol observations to provide feedback on a questionnaire from the European Radio Spectrum Policy Group (RSPG), which is seeking to coordinate the use of the 6-GHz band between IMT, Wi-Fi and other stakeholders, including the RAS in the European Union. Finally, CRAF encourages a more direct interaction between CRAF and scientists using VLBI as well as with the staff at the correlators to discuss and address RFI-related issues. The outcomes of these discussions could then serve as important and concrete examples and a foundation in the spectrum management arena.

\begin{acknowledgements}
The authors like to thank all CRAF delegates, observer members and collaborators as the work presented in this contribution stems from a community effort.
\end{acknowledgements}

\bibliographystyle{aa}
\bibliography{references}

\end{document}